\documentstyle[titlepage,fleqn,12pt]{article}
\begin{document}

\begin{center}
{\Large \bf{Structural Transition Models for a\\
\vspace{0.5cm}
class of Irreversible Aggregates}} \\
\vspace{2cm}
{\bf Enrique Canessa}\footnote{E-mail: canessae@ictp.trieste.it} \\
{\em ICTP-International Centre for Theoretical Physics,}
{\em Condensed Matter Group} \\
{\em P.O. Box 586, 34100 Trieste, ITALY}
\end{center}

\vspace{2cm}
{\baselineskip=20pt
\begin{center}
{\bf Abstract}
\end{center}
A progress report on two recent theoretical approaches proposed to
understand the physics of irreversible fractal aggregates showing up a
structural transition from a rather dense to a more multibranched growth
is presented.  In the first approach the transition is understood by
solving the Poisson equation on a squared lattice.  The second approach is
based on the discretization of the Biharmonic equation.
Within these models the transition appears when the growth
velocity at the fractal surface presents a minimum.
The effects of the surrounding medium and geometrical
constraints for the seed particles are considered.  By using the optical
diffraction method, the structural transition is further characterized by
a decrease in the fractal dimension for this peculiar class of aggregates. \\


\vspace{2.5cm}
}

\baselineskip=22pt
\parskip=4pt

\pagebreak

\section{Introduction}

Some irreversible fractal aggregates display in nature peculiar structural
transitions during their growing processes.  Electrochemical deposition
experiments \cite{Fle92} and bacterial colony growth \cite{Ben92} are two
examples of such phenomena.  It has been reported that, at a certain threshold
distance far from the cluster center, these systems exhibit an intriguing
transition from a rather dense to a more multibranched growth of which little
is understood.

Besides the classical diffusion limited aggregation for 'regular' fractals
(having fractal dimension $d_{f}\sim 1.7$) \cite{Wit83,Ziq95}, the Laplacian
aggregation model has been proposed which is based on the discretized Laplace
equation\cite{Nie84}
\begin{eqnarray}\label{eq:l1}
 \phi_{i,j} &=& \frac{1}{4}(\phi_{i+1,j}+\phi_{i-1,j}+\phi_{i,j+1}+
  \phi_{i,j-1}) \;\;\; ,
\end{eqnarray}
with $\phi_{i,j}$ the potential growth site.  However, neither of these simple
approaches lead to explain a complex structural transition during irreversible
fractal growth.

Two alternative attempts have recently been reported in the literature to
understand the physics behind a structural transition in fractal growth.
In the first approach, due to Louis {\em et al.}
\cite{Lou92,Cas93,Wan93a}, the transition is derived by solving
the Poisson equation (on a squared lattice)
\begin{equation}\label{eq:w0}
\nabla^{2}\phi = \lambda^{2}\phi \;\;\; ,
\end{equation}
which becomes dependent on the potential $\phi$ at two boundaries, the distance
$L$ between them, and a screening length $\lambda$.

The second theoretical approach is due to Wang and Canessa \cite{Wan93b,Can93}
and is based on the Biharmonic equation in two-dimensional (2D)
isotropic defect-free media, namely
\begin{equation}\label{eq:w1}
\nabla^{2} (\nabla^{2} u)=0   \;\;\; .
\end{equation}
By discretizing this equation, these authors have shown that a structural
transition can also be a consequence of the different
coupling of displacements $u$ within the pattern formation.

Both of these approaches describe a similar class of complex topology
from different perspectives and on different systems.  However, the Poisson
and Biharmonic growth models identically relate the growth probability for
each site to the local field as
\begin{eqnarray}\label{eq:w2}
P \approx \phi &  \;\;\; or \;\;\; &
P \approx \nabla ^{2}u \;\;\; ,
\end{eqnarray}
These assumptions phenomenologically follow the Laplacian
(or dielectric breakdown) model.

It is the aim of this progress report to review the Poisson and Biharmonic
approaches.  We shall focus on these two model for irreversible fractal
aggregates showing up a structural transition and shall consider the
effects of the surrounding medium
and geometrical constraints for the seed particles.  Following recent ideas by
P\'erez-Rodr\'iguez {\em et al.} \cite{Per94}, we shall also shown how the
optical diffraction method leads to characterize the structural
transition by a decrease in the fractal dimension for this class of aggregates.

\section{The Poisson Growth Model}

The origin of screening in the Poisson model lies in the presence of free
charges and leads to a rich variety of patterns \cite{Lou92}.  As compared to
the Laplacian model of Eq.(\ref{eq:l1}), this model introduces a new length
scale, {\em i.e.}, $\lambda$, and a nontrivial dependence on the boundary
conditions that are responsible for a structural transition on growing.

Fig.1(a) shows an example of the Poisson model in planar geometry by
adding 4051 particles to the cluster and using $\phi^{i}=5 \times 10^{-7}$,
$\phi^{o}=1$ and $\lambda^{-1}=10$.
{}From this figure it can be seen that the Poisson patterns obtained can have a
fractal character (at scales shorter than $\lambda$), to then display a
structural transition.

Louis {\em at.} \cite{Lou92} have analytically demostrated that this
structural transition can be characterized by a change in the sign of
the electrostatic field at the surface of the aggregate besides the
minimum of the growth velocity (to be shown later).  However, Wan Wei
\cite{Wan93a} has shown that such a transition may be
altered by the existence of a critical field.

It is important to note that within the Poisson model several fractal
branch might also grow when attaching more than one particle at each
computer step.  For details see Refs.\cite{Lou92,Cas93}.

\section{The Biharmonic Growth Model}

Fig.1(b) shows the final stage of a Biharmonic fractal displaying features
(in circular geometry) of a transition.   Below the transition point
$r_{\ell}$, the fractal dimension approaches
the value for Laplacian growth within error bars.  To generate this class of
patterns one solves numerically the Biharmonic Eq.(\ref{eq:w1}).  Its discrete
form on the ($i,j$) lattice site, yields the following expression
\begin{eqnarray}\label{eq:w4}
& & u_{i-2,j}+2u_{i-1,j-1}-8u_{i-1,j}+2u_{i-1,j+1}+
  u_{i,j-2}-8u_{i,j-1}+20u_{i,j}  \\ \nonumber
  & & -8u_{i,j+1}+u_{i,j+2}+
2u_{i+1,j-1}-8u_{i+1,j}+2u_{i+1,j+1}+u_{i+2,j}=0   \;\;\;
\end{eqnarray}
requiring the values of the normal derivative for the order parameter $u$
\cite{Gde86}.  For the sake of simplicity one sets the derivative boundary
condition (necessary along the radial direction) equal to zero.

Throughout calculations one uses lattice sites enclosed within a circle of
(normalized) radius $r=\sqrt{i^{2}+j^{2}}$ such that $u^{o}$ and $u^{i}$ are
unity and zero at the outer circular boundary and the inner growing Biharmonic
aggregate, respectively.  The seed particle can be placed either centered or
distributed under different geometrical constraints as discussed later.
The procedure for growing fractals then follows standard techniques
\cite{Nie84}, till solutions of the discretization of Eq.(\ref{eq:w1})
converge. Aggregates then stochastically grow under
the relation between the growth probability $P$ (at the grid site $(i,j)$) and
$u$ as in Eq.(\ref{eq:w2}).

\section{Growth Velocity}

Results for the grow velocity $v$ along the $y$-direction for the Biharmonic
model are plotted in Fig.2.  For comparison, also included in this figure are
the results for Poisson growth assuming $v$ to be proportional to the field
$\mid \phi_{i,j}-\phi^{i}\mid$ by following Ref.\cite{Lou92}.

{}From Fig.2, it can be seen that the structural transition coincides
with the fact that $v$ on the growing surface presents a minimum.
The same is true for Biharmonic growth (even independently of how one relates
the probability $P_{ij}$ to $u(i,j)$ \cite{Wan93b}).  For Laplacian
growth this phenomenon does not appear because the trend is to generate a
single tip at faster velocity than in the cases of Poisson or Biharmonic
growth.

Biharmonic patterns below $r_{\ell}$, ({\em i.e.}, within the dense region of
Fig.1(b)), does not become that dense as for Poisson of Fig.1(a).  This
effect can be understood from the velocity plot since for Poisson growth
the transition occurs at smaller velocities than for Biharmonic
growth.  Henceforth, a Eden-like pattern can be generated due to screening.
Both Poisson and Biharmonic models lead to growth velocities along the
$y$-direction with parallel slopes.

\section{Effects of Surrounding Medium and Geometrical Constraints}

Within the Biharmonic model, the effects on fractal growth of the surrounding
medium and of different geometrical constraints (or boundary conditions) for
the seed particles has been analysed in Ref.\cite{Can93} by following
the $\eta$ (or dielectric breakdown) model \cite{Nie84,Mea91} and assuming
an extention of Eq.(\ref{eq:w2}) as
\begin{equation}\label{eq:w22}
P_{ij}= \frac{\mid \nabla ^{2}u_{i,j}\mid^{\eta} }{\sum \mid \nabla^{2}
        u_{ij}\mid^{\eta} }  \;\;\; ,
\end{equation}
where the sum runs over nearest neighbor sites and $\eta \ge 0$.

Figure 3(a) shows typical numerical results by using $\eta =2$ for a cluster
of about 1000 particles.  In the limit $\eta \rightarrow \infty$, a
{\em 'transition from slow to faster growth'} is found
in such a way that one end of the
needle-like structure presents greater growth probability than the other.
Such structures become dendritic below and above a transition point.
On decreasing the value of $\eta$, one finds a transition from
{\em dendritic-to-compact} growth such that the inner region of
the aggregates becomes denser.  If $\eta \rightarrow 0$, the growth
probability becomes purely random and independent of the Biharmonic field
-as in a Eden-like pattern.

The effects of geometrical constraints for the seed particles has also been
studied in Ref.\cite{Can93} by fixing the value of $\eta$ and using
two different geometries for the seed particles, namely circular and linear.
Figure 3(b) shows a Biharmonic fractal under the constrains of a circular
(empty) area of seed particles.  As can be seen
a structural transition still survives independently of such a
simple configuration adopted.

\section{Optical Diffraction Analysis}

The effect of the structural transition on the diffracted intensity and,
consequently, on $d_{f}$ follows by considering the
fractal to be composed of $N$ identical and similarly oriented particles on
the plane $x-y$ \cite{Ber91,Kor92}.  The position of their centers of mass
is given by
${\bf R}_{n}=(x_{n},y_{n})$, where $n=1,...,N$. In this system, the Fraunhofer
diffraction pattern for the fractal is derived by assuming that each particle
corresponds to one aperture.

The form factor, corresponding to the intensity scattered by one 'aperture',
is determined by the integral of the diffraction amplitude.  The structure
factor is
\begin{equation}\label{eq:a4}
S({\bf k})\equiv \mid \frac {1}{N} \sum _{n=1}^{N}
e^{-i{\bf k}{\bf R}_{n}}\mid ^{2} \;\;\; ,
\end{equation}
where {\bf k} is the component, parallel to the $xy$-plane, of the
scattered wave vector whose modulus is $k = \frac {2\pi }{\lambda }
{\rm sin}\theta \approx \frac {2\pi }{\lambda } \theta$, with $\theta $ the
(small) angle that the scattered wave vector makes with the $z$-axis.
$\lambda $ is the wavelength of the incident light.  If $ka<1$ ($a$ being the
size of an elementary particle), the form factor is practically unity and the
light distribution in the diffraction pattern is given by the structure factor
such that the normalized diffraction intensity is
\begin{equation}
I({\bf k})\approx S({\bf k}) \;\;\; .
\end{equation}

For $\frac{a}{L}<ka<1$ ($L$ being the size of the whole aggregate) the expected
value of the intensity -or, alternatively, of the structure
factor $S({\bf k})$- is
\begin{equation}\label{eq:lq2}
<I({\bf k})>= \int d^{2}{\bf R} e^{-i{\bf k}{\bf R}}
 ( \int d^{2}{\bf R}_{0} \rho ({\bf R}_{0})
\rho ({\bf R}+{\bf R}_{0}) \; \sim R^{-\alpha} )\;\;\; .
\end{equation}
This relation for fractal aggregates obeys a power law behaviour \cite{Wit83},
where the exponent $\alpha$ is related to the Hausdorff dimension
$d_{f}=d-\alpha$ with $d$ the Euclidean space dimension.  This variation, in
turn, leads to the power law behavior of the diffracted intensity as a function
of the wave vector
\begin{equation}
I({\bf k})\sim k^{-d_{f}} \;\;\; .
\end{equation}

Information about $d_{f}$ can be obtained at low {\bf k} values (such that
$\frac{a}{L} \ll \mid k_{x}a \mid \ll 1$) from the averaged intensity defined
by the expression
\begin{equation}\label{eq:a16}
<I(k)>\equiv \frac {1}{2\pi} \oint d\phi I({\bf k}) \ \ ;
\ \ \ k_{x}=k{\rm cos}\phi, \ \ k_{y}=k{\rm sin}\phi,
\end{equation}
As seen in Fig.4, this quantity for Poisson fractals has a power law behavior
as a function of the modulus $k$. However, the structural transition during
fractal growth leads to a change in $d_{f}$.  The slopes of this plot represent
a decrease of the fractal dimension for Poisson aggregates.  Similar results
have been found in Biharmonic fractals \cite{Per94}.

\section{Discussion}

In both models studied the global influence of a growing pattern to the growth
probability for each lattice site under a certain power-law form is set
phenomenologically as in Eq.(\ref{eq:w2}) or (\ref{eq:w22}).
However an important difference to note
is that within the Biharmonic model, iterative procedures are carried
out around {\em thirteen} next nearest neighbors and not on {\em four} as
within the Poisson and Laplacian models.  Thus the formation of connected
patterns within the Biharmonic equation becomes non trivial and more involved.

Within the Poisson model, screening lies in the presence of free charges.
The physical relevance for the Biharmonic equation might follow from elasticity
theory (the deflection of a thin plate subjected to uniform loading over its
surface with fixed edges), the steady slow 2D motion of a viscous
fluid or the vibration modes in the acoustic of drums.  Besides this, a higher
order differential equation containing the Biharmonic term also appears in the
study of kinetic growth with surface relaxation (see, {\em e.g.}, recent work
in \cite{Yan92} and references therein).

The influence of ramified Biharmonic patterns on the growth probability
for each lattice site is on the type of pattern obtained and not on locating
the transition.  Several branches may also develop within the Poisson and
Biharmonic growth models by attaching -simultaneously, and stochastically-
more than one particle at each time step.  For Poisson growth the structural
transition is the result of including many-body contributions via screening
in a sort of mean field approach.  In the Biharmonic model, long range
coupling appears naturally as a consequence of discretizing Eq.(\ref{eq:w1})
(as given in Eq.(\ref{eq:w4})).
For planar symmetry, the transition point within both models appears when the
growth velocities exhibit a minimum. This implies that Poisson and Biharmonic
growth describe a similar class of complex structural transition phenonema
from two different perspectives and on different systems.

The structural transition point can, to a good approximation, be estimated
from the continuous limit of Eq.(\ref{eq:w1}) in cylindrical coordinates.
Canessa and Wang \cite{Can93} have shown that the transition point depends on
the system size $L$ only and occurs approximately at a distance (in lattice
units) about $60 \%$ far from the seed particle indicated by the circle arrows
in Fig.1(b).  This prediction is in accord with their numerical
simulations for $\eta =1$.

By tuning $\eta \rightarrow 0$ in Eq.(\ref{eq:w22}), the structural transition
corresponds to a {\em 'dense-to-multibranched transition'} whereas
for $\eta \rightarrow \infty$ one founds a {\em 'transition from slow to
faster growth'}.  On decreasing
$\eta$ from infinity to zero, Biharmonic fractals become denser as for
Laplacian growth.  For large $\eta$, the needle-like structures grow faster
at one end and, in the overall, remain denditric.  Furthermore the transition
point is independent of geometrical constrains adopted for the seed particles.

Results of optical diffraction have enabled to identify and relate changes
in $d_{f}$ of aggregates to variations in the diffracted intensity as a
function of the wave vector.  Such findings may be experimentally confirmed.
In fact, the averaged intensity
$<I(k)>$ might be determined by using the set up as the one described in
\cite{All86}.  Therein, a photomultiplier connected to a multichannel
analyzer records $I(k_{x},k_{y})$ and the displacement of the photomultiplier
is controlled by a high-precision motorized micrometer. Then, after
scanning the diffraction patterns displaying a structural transition, the
average of intensity over concentric circles ($<I(k)>$) might be obtained.

\begin{center}
{\bf  Acknowledgments}
\end{center}

The author would like to thank Dr Wang Wei (Najing University, China) and
Dr Felipe P\'erez-Rodr\'iguez (Universidad Autonoma de Puebla, M\'exico) for
a fruitful collaboration on the present topic.  The Scientific Computer
Section and the Condensed Matter Group at ICTP-Trieste, Italy, are also
acknowledged for financial support.

\newpage

\section*{Figure captions}

\begin{itemize}
\item {\bf Fig.1}:  Different types of irreversible aggregates displaying
the transition.  (a) Poisson fractal in planar geometry. (b) Biharmonic
fractal in circular geometry.

\item {\bf Fig.2}: Growth velocity $v$ along one direction for two
different Poisson fractals (grouped as I)
and for three different Biharmonic fractals (grouped as II).

\item {\bf Fig.3}: (a) Biharmonic fractal in circular geometry using in
Eq.(\ref{eq:w22}) $\eta =2$ to mimic surrounding medium effects.
(b)Biharmonic fractal under the constrains of a circular
(empty) area of seed particles.

\item {\bf Fig.4}: Log-Log plot of the angle-averaged $<I(k)>$ against $k$
for a typical Poisson fractal.  Full and open squares, represent results
for this class of cluster before and after the structural transition.

\end{itemize}


\newpage

\begin{thebibliography}{99}

\bibitem{Fle92} V. Fleury, J.-N. Chazalviel and M. Rosso, Phys. Rev.
Lett. {\bf 68}, 2942 (1992).
\bibitem{Ben92} E. Ben-Jacob, H. Shmueli, O. Shochet and A. Tenenbaum,
Physica A {\bf 187}, 378 (1992).
\bibitem{Wit83} T.A. Witten and L.M. Sander, Phys. Rev. B {\bf 27},
5686 (1983).
\bibitem{Ziq95} W. Ziquin and L. Boquan, Phys. Rev. B {\bf 51}, R16 (1995).
\bibitem{Nie84} L. Niemeyer, L. Pietronero and H.J. Wiesmann, Phys. Rev. Lett.
{\bf 52}, 1033 (1984).
\bibitem{Lou92} E. Louis, F. Guinea, O. Pla and L.M. Sander,
Phys. Rev. Lett. {\bf 68}, 209 (1992).
\bibitem{Cas93} J. Castella, E. Louis, F. Guinea, O. Pla and
L.M. Sander, Phys. Rev. E {\bf 47}, 2729 (1993).
\bibitem{Wan93a} W. Wang, Phys. Rev. E {\bf 47}, 2893 (1993).
\bibitem{Wan93b} W. Wang and E. Canessa, Phys. Rev. E {\bf 47}, 1243 (1993).
\bibitem{Can93} E. Canessa and W. Wang, Z. Naturforsch. {\bf 48a}, 945 (1993).
\bibitem{Per94} F. P\'erez-Rodr\'iguez, W. Wang and E. Canessa, Opt. Comm.
{\bf 108}, 185 (1994).
\bibitem{Yan92} H. Yan, Phys. Rev. Lett. {\bf 68} 3048 (1992).
\bibitem{Gde86}  G. de Vahl Davis, {\em "Numerical Methods in Engineering
and Science"} (Allen $\&$ Unwin Ltd., London, 1986).
\bibitem{Mea91}  P. Meakin, J. Feder and T. J$\phi$ssang, Phys. Rev. A.
{\bf 43}, 1952 (1991).
\bibitem{Ber91} D. Berger, S. Chamaly, M. Perreau, D. Mercier, P. Monceau and
J.C.S. Levy, J. Phys. I (France) {\bf 1}, 1433 (1991).
\bibitem{Kor92} G. Korvin, in {\em 'Fractal Models in the Earth Sciences'},
(Elsevier, Amsterdam, 1992).
\bibitem{All86} C. Allain and M. Cloitre, Phys. Rev. B {\bf 33}, 3566 (1986).

\end{thebibliography}
\end{document}